\documentclass[twocolumn,prb,amsmath,amssymb,amsfonts,superscriptaddress,floatfix,showpacs]{revtex4}

\usepackage{graphicx}
\usepackage{dcolumn}
\usepackage{bm}
\usepackage{rotating} 

\begin{document}

\title{Electronic structure of rare-earth impurities in GaAs and GaN}
\author{A. Svane}
\affiliation{Department of Physics and Astronomy, University of Aarhus, DK-8000 Aarhus C, Denmark}
\author{N. E. Christensen}
\affiliation{Department of Physics and Astronomy, University of Aarhus, DK-8000 Aarhus C, Denmark}
\author{L. Petit}
\affiliation{Computer Science and Mathematics Division, and Center for Computational 
 Sciences, Oak Ridge National Laboratory, Oak Ridge, TN 37831, USA}
\author{Z. Szotek}
\affiliation{Daresbury Laboratory, Daresbury, Warrington WA4 4AD, UK}
\author{W. M. Temmerman}
\affiliation{Daresbury Laboratory, Daresbury, Warrington WA4 4AD, UK}

\date{\today}

\begin{abstract}
The electronic structures of substitutional rare-earth (RE) impurities in GaAs and cubic GaN
are calculated.  The total energy is evaluated with the
self-interaction corrected local spin density approximation, by which several configurations
of the open 4f shell of the rare-earth ion may be investigated. The defects are modelled by supercells
of type REGa$_{n-1}$As$_n$, for $n=4$, $ 8$ and $16$. The preferred defect is the rare-earth
substituting Ga, for which case the rare-earth  valency 
in intrinsic material is found to be
trivalent in all cases except Ce and Pr in GaN.
The $3+$ $\rightarrow$ $2+$ $f$-level is found above the theoretical conduction band edge in all cases and
within the experimental gap only for Eu, Tm  and Yb in GaAs and for
Eu in GaN.
The exchange interaction of the rare-earth impurity with the states at both the
valence band maximum and the conduction band minimum
is weak, one to two orders of magnitude smaller than that of Mn impurities. Hence the
 coupling strength
is insufficient to allow for ferromagnetic ordering of dilute impurities, except at very low temperatures.
\end{abstract}

\maketitle

\section{Introduction}

The technology of electronics exploiting both charge- and spin-degrees
of freedom of carriers is a rapidly developing research field.\cite{wolf,zutic}
Diluted magnetic semiconductors (DMS) are good candidates for new generations
of fast, low dissipation, non-volatile integrated information and processing
devices. In particular, ferromagnetic semiconductors offer a 
unique combination of magnetic, semiconducting
and optical properties, together with  compatibility with existing microelectronics
technology. In the longer terms, the long electron lifetimes in semiconductors
coupled with their strong interplay with nuclear spins might be exploited for
quantum computing.\cite{kawakami}
Several reports of ferromagnetically ordered DMS at room temperature 
(see {\it e.g.} Refs. \onlinecite{liu,macdonald} and references therein) 
pose  questions of fundamental importance as to the nature of magnetic interactions in 
these materials.\cite{dietl,jungwirth,zunger,dederichs,sharma,NatMat,leon}

Most efforts have investigated Mn and other $3d$ impurities
in III-V and II-VI hosts, but recently also the 
discovery of high temperature ferromagnetism\cite{teraguchi,ploog}
in diluted Gd-doped GaN has gained considerable interest, implying that rare-earth (RE) ions may
become a viable alternative to transition metals for spintronics applications.
In addition, rare-earth doping of wide-gap semiconductors such as GaN and ZnO, 
 is an active research area in itself due to
the possibility of achieving tunable light emission,\cite{steckl} arising from 
intra-$f$-shell optical transitions.

The present work undertakes a systematic theoretical investigation of the electronic 
structure of rare-earth dopants in GaAs and GaN, both in the zinc-blende crystal
structure.
Total energy calculations reveal Ga substitution as the most favorable rare-earth 
defect, with the rare-earth ion taking the iso-electronic trivalent configuration in all
cases studied except Ce and Pr in GaN. The effect of the
exchange interaction of the rare-earth ion with the host states at
the valence band maximum (VBM) and conduction band minimum (CBM)
is calculated, and is found surprisingly weak given the large spin moment formed on the
rare-earth ion. A significant hybridization effect between the unoccupied $f$-states and the 
CBM states is observed. Furthermore, the position of the divalent acceptor level
$\epsilon(0/-)$ is
calculated and found outside the theoretical band gap, {\it i.e.} 
appearing as a resonance in the conduction bands. However, 
the $\epsilon(0/-)$ level falls
inside the experimental
band gap for Eu in GaN and for Eu, Tm and Yb in GaAs.

The theoretical modelling   of solids containing rare-earths poses severe problems due
to the correlated nature of the $f$ electrons in the incompletely filled $4f$ shell. 
A description in terms of bands, as implemented in the conventional 
local spin density (LSD) approximation\cite{gunnarsson,martin} invoked in the framework
of density-functional theory,\cite{kohn} leads to a large overestimation of the
contribution of $f$-electrons to bonding and hence too small lattice constants.
In addition, even if  the band structures of LSD have no formal interpretation, they are often
used as guidelines for what the elementary electron excitations of the solid are, but in the case of
$f$-electron materials the approach is particularly poor.    The
$f$ derived states appear as a set of narrow bands situated
close to  the Fermi level, 
while experiments reveal their localized nature as
atomic multiplet features over a broad energy range.\cite{Campagna}

In the present work the electronic structure of rare-earth impurities in semiconductors is    
 investigated with the 
self-interaction corrected (SIC) local spin density method, which constitutes an extension of
the conventional LSD approximation
 by correcting          for the spurious
self-interaction of individual electrons.\cite{zp} For rare-earth atoms the self-interaction correction
allows for a localization of the $f$-electrons. In this
way the atomic limit of the rare-earth $f$ shell
is     implemented in conjunction with the itinerant limit of the other
rare-earth valence electrons and        the valence electrons of the semiconducting
host. In particular, the ground state configuration of the rare-earth $f$-shell is determined 
from comparison of the total energy corresponding to different valency scenarios,
as has already been shown in several applications to rare-earth solids,\cite{nature,yb,ybco}
including pressure induced valence transitions.\cite{ce,cep,sms}

In section II of the present paper we outline the theoretical aspects of the present work, notably the
SIC-LSD total energy method as well as other calculational details. In section III the results are presented
for defect formation energies, magnetic interactions, and valency transition energies. Section IV 
contains the summary and conclusions of this work.

\section{Theory}

\subsection{The SIC-LSD total energy method}

The total energy functional of the
LSD approximation is renowned for its chemical accuracy
in describing conventional solids with weakly correlated electrons.\cite{gunnarsson,martin} 
The self-interaction correction is included
to facilitate an accurate description of the localized $f$ electrons of
 rare-earths.
Specifically, the SIC-LSD \cite{zp} total energy functional is obtained from the LSD
as:
\begin{eqnarray}
\label{Esic}
E^{SIC-LSD}&=&E^{LSD}+\Delta E_{sic}+E_{so},\\
\Delta E_{sic}&=&-\sum_{\alpha }^{occ.}\delta _{\alpha }^{SIC}, \\
E_{so}&=&\sum_{\alpha }^{occ.}\epsilon_{\alpha}^{so}
\end{eqnarray}
where $\alpha $ labels the occupied states and $\delta _{\alpha }^{SIC}$ is
the self-interaction correction for state $\alpha $. As usual, $E^{LSD}$ can
be decomposed into a kinetic energy, $T$, a Hartree energy, $U$, the
interaction energy with the atomic ions, $V_{ext}$, and the exchange and
correlation energy, $E_{xc}$.\cite{kohn} 
The self-interaction is defined as
the sum of the Hartree interaction and the exchange-correlation energy for
the charge density of state $\alpha $: 
\begin{equation}
\label{dsic}
\delta _{\alpha }^{SIC}=U[n_{\alpha }]+E_{xc}[n_{\alpha }].
\end{equation}
For itinerant states, $\delta _{\alpha }^{SIC}$ vanishes identically, while
for localized (atomic-like) states the self-interaction may be appreciable.
This correction constitutes a negative energy
contribution for an $f$-electron to localize, 
which then competes with the band formation energy gained by the
$f$-electron if allowed to delocalize and hybridize with the available conduction states. 
In rare-earths the self-interaction correction
 ranges from $\delta_{\alpha}\sim 0.8$ eV  per $f$-electron in Ce to
$\delta_{\alpha}\sim 1.5$ eV  per $f$-electron in Yb, reflecting the contraction of the $f$-orbitals
through the series.
The volume dependence of $\delta_{\alpha}$ is rather weak, hence
  the overbinding 
of the LSD approximation for narrow band states is reduced when localization is allowed.
The last term in Eq. (1) is the spin-orbit energy, where for each occupied state $\alpha$:
\begin{equation}
\label{Eso}
\epsilon_{\alpha}^{so}= \langle \psi_{\alpha} | \xi(\vec{r})\vec{l}\cdot\vec{s} | \psi_{\alpha} \rangle 
\end{equation}
is added to the energy functional.
We employ the atomic spheres approximation, whereby the crystal volume is divided into 
slightly overlapping atom-centered spheres of a total volume equal to the actual volume.
In (\ref{Eso}), the angular momentum operator, $\vec{l}=\vec{r}\times\vec{p}$, 
is defined inside each atomic sphere, with $\vec{r}$ given as the position vector
from the sphere center.
The spin-orbit interaction couples the band Hamiltonian for the spin-up and spin-down channels, 
i.e.  a doubled secular problem must     be solved.
Other relativistic effects are automatically included by solving the scalar-relativistic
radial equation inside spheres.
The spin-orbit parameter, 
\[
\xi(r)=-\frac{2}{c^2}\frac{dV}{dr}
\]
 in atomic Rydberg units, 
is calculated from the self-consistent potential.
The SIC-LSD energy functional in Eq. (1) appears to be a functional of all the one-electron orbitals,
but can in fact be viewed as a functional of the total (spin) density alone, as discussed in Ref.
\onlinecite{comment}.

The advantage of the SIC-LSD energy functional 
is that it allows for different valency scenarios to  be
explored. By assuming atomic configurations with different total numbers of
localized states, self-consistent minimization of the total energy leads to different local
minima of the same functional, $E^{SIC-LSD}$ in Eq. (1), 
and hence their total energies may be compared. The
configuration with the lowest energy defines the ground state configuration. Note,
that if no localized states are assumed, $E^{SIC-LSD}$ coincides with the
conventional LSD functional, i.e., the Kohn-Sham minimum of the $E^{LSD}$
functional is also a local minimum of $E^{SIC-LSD}$. The interesting question
is, whether competing minima with a finite number of localized states exist.
This is usually the case in $f$-electron systems \cite{nature}
 and some 3d transition
metal compounds,\cite{tmo} where the respective $f$ and $d$
orbitals are sufficiently confined in space to benefit appreciably from the
SIC.

The SIC-LSD still considers the electronic structure of the solid to be
built from individual one-electron states, but offers an alternative description
to the Bloch picture, namely in terms of
periodic arrays of localized atom-centered states ({\it i.e.}, the
Heitler-London picture in terms of Wannier orbitals). Nevertheless, there
still exist states which will never benefit from the SIC. These states
retain their itinerant character of the Bloch form, and move in the
effective LSD potential. This is the case for the host valence band states in the systems studied here.
The SIC-term only affects the rare-earth ions. The resulting many-electron wavefunction will 
consist of both localized and itinerant states. 
The nominal valency of a rare-earth ion is defined as the number of valence electrons available for band
formation:
\[
N_{val}=Z-N_{core}-N_{SIC},
\]
where $Z$ is the nuclear charge, $N_{core}$ the number of core electrons, and $N_{SIC}$ the number of
self-interaction corrected $f$-electrons.

In
contrast to the LSD Kohn-Sham equations, the SIC electron states, which minimize
$E^{SIC-LSD}$, experience different effective potentials. This implies
that to minimize $E^{SIC-LSD}$, it is necessary to explicitly ensure the
orthonormality of the one-electron wavefunctions by introducing a Lagrangian
multipliers matrix. Furthermore, the total energy is not anymore invariant
with respect to a unitary transformation of the one-electron wavefunctions.
Both of these aspects make the energy minimization more demanding to
accomplish than in the LSD case. The electron wavefunctions are expanded in 
linear-muffin-tin-orbital (LMTO) basis functions,\cite{OKA}
 and the energy minimization problem becomes a
non-linear optimization problem in the expansion coefficients.
Further details of the present implementation
can be found in Ref. \onlinecite{brisbane}.

   \subsection{Calculational Details} 
Alloys of GaAs (or GaN) and rare-earth ions of the type  RE$_x$Ga$_{1-x}$As 
(RE$_x$Ga$_{1-x}$N) and 
RE$_x$GaAs$_{1-x}$ (RE$_x$GaN$_{1-x}$) were considered by supercell modelling, and
total energies evaluated for $x=1/4,$ $1/8$ and $1/16$. The experimental lattice constants of 
pure GaAs and GaN in the zinc-blende structure were used ($a=5.65$ \AA\ for GaAs,
$a=4.45$ \AA\ for GaN), {\it i.e.} 
the effects of lattice expansion with doping were assumed negligible.
The atomic positions were kept fixed in all calculations, except for a few test cases discussed in the next
section, {\it i.e.} no systematic study of the effects of relaxations of nearest neighbours
of the impurities were attempted. Experimentally, for Er-doped GaN the Er-N distance has been found 
$\sim 10 \%$ larger than the pure host Ga-N bond length.\cite{citrin}

The itinerant states were sampled using 95, 22 and 
14 k-points in the irreducible wedge of the Brillouin zone of 
the $x=1/4,$ $1/8$ and $1/16$ supercell, respectively.
In GaAs, the $3d$ semicore states of Ga and As were treated selfconsistently in a separate energy panel,
which also included the $5s$ and $5p$ semicore states of the rare-earth.
In the case of the GaN host, the best description is obtained with
only the $5s$ semicore states of the rare-earth impurity treated
in the separate semicore panel. 
Due to the smaller lattice constant, there is a pronounced interaction between the N $2s$, $2p$ states and 
the Ga $3d$ and rare-earth $5p$ states, for which reason the latter were treated as valence electrons
to describe properly their hybridization into the GaN valence bands.

\section{Results and Discussion}

   \subsection{Cohesive Properties}
\begin{figure}
\begin{center}
\includegraphics[width=90mm,clip]{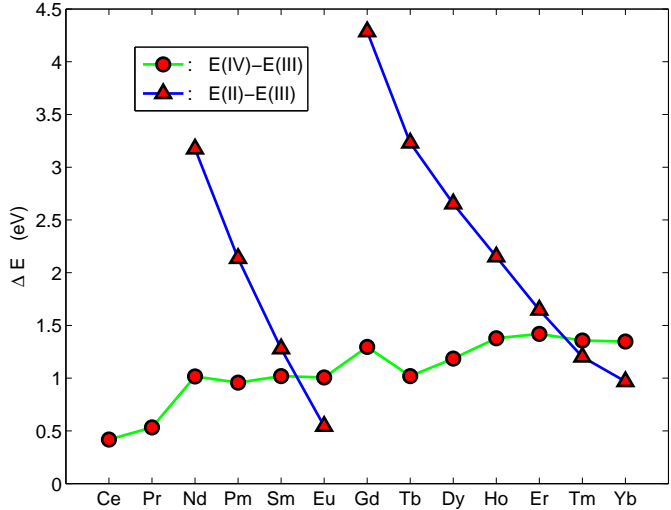}\\
\caption{ 
\label{GaAs_val}
(Color online)
The energy difference between tetravalent and trivalent rare-earth impurities substituting for
Ga in GaAs (circles), and for selected cases between divalent and trivalent ions (triangles). The calculations
were performed for supercells corresponding to rare-earth dilution
$x=1/16$. The positive sign implies that the trivalent configuration
is the ground state.
}
\end{center}
\end{figure}

\begin{figure}
\begin{center}
\includegraphics[width=90mm,clip]{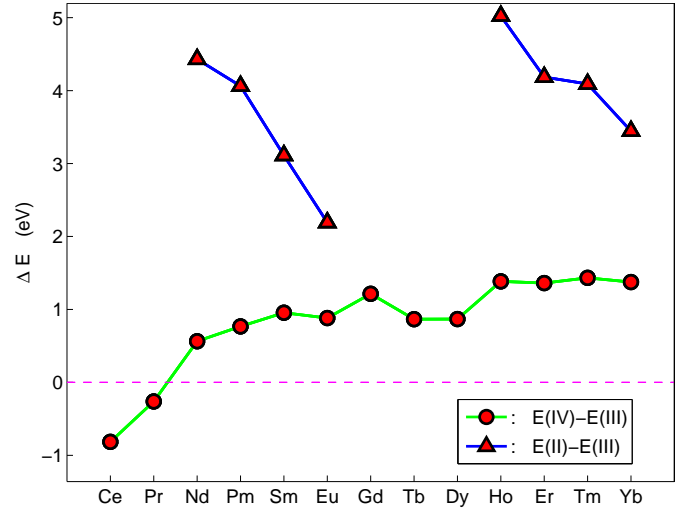}\\
\caption{ 
\label{GaN_val}
(Color online)
The energy difference between tetravalent and trivalent rare-earth impurities substituting for
Ga in cubic GaN (circles), and for selected cases between divalent and trivalent ions (triangles). The calculations
were performed for supercells corresponding to rare-earth dilution
$x=1/16$.
}
\end{center}
\end{figure}

Figures \ref{GaAs_val} and \ref{GaN_val} show the energy differences 
calculated for trivalent, divalent and tetravalent configurations of the rare-earth
impurities, in GaAs and GaN, respectively, at concentration $x=1/16$. 
The rare-earth ion is taken   to substitute for Ga. 
The preferred valency is seen to be trivalent for all rare-earth impurities in GaAs, and for all
but Ce and Pr in GaN, in good accord with the trivalency of the Ga atom which the rare-earth
replaces.  In particular, the Eu and Yb impurities remain trivalent (by a margin of more than 0.5 eV)
in spite of the tendency of these elements to become divalent, {\it e.g.} in the elemental metallic
phase.
Luminescence experiments show that Eu in GaN is trivalent,\cite{nyein} although
there have been indications of divalent Eu in the surface region of Eu-doped 
GaN.\cite{Maruyama}

In GaAs, the valence energy difference is almost independent of concentration, except for Ce, 
for which case,
at concentrations $x=1/4$ and $x=1/8$ the
 preferred configuration is $f^0$ (tetravalent). 
This  means that $f$-band formation is favored for Ce impurities at this
large concentration, i.e. instead of having one localized
$f$-electron on each Ce atom, a narrow $f$-band is formed close to the conduction band minimum, and filled
with one electron leading to a metallic alloy.
However, in the most diluted case studied here, $x=1/16$, the localization of one $f$-electron on each Ce
is more favorable, leading to a ground state with cerium ions in the trivalent $f^1$ configuration 
and a semiconducting alloy, which we hence assume to
prevail also at Ce concentrations of even lower value.

In GaN both Ce
and Pr are found to be tetravalent, even at $x=1/16$, i.e. again implying the formation of
narrow $f$-bands straddling the conduction band edge and metallicity. The band formation is favored by  the 
small host lattice constant, which particularly affects the least tightly bound $f$-electrons in the early
lanthanides.  Presumably, further dilution beyond what has been studied here, will eventually 
lead to the formation of localized trivalent ions also for Ce and Pr in GaN. However, also
effects of relaxations of the nearby host atoms are important. As a check, a similar valency energy
difference was calculated for Ce in GaN with the nearest neighbor N atoms moved 10\% 
of the bond length away from Ce, and indeed
the trivalent configuration in this case had a lower energy (by 0.14 eV) than the tetravalent configuration.
The detailed analysis of the
localization-delocalization transition upon dilution would need to incorporate besides the relaxation of 
atomic positions in the vicinity of the impurity, also issues of impurity disorder and/or clustering,
which is not attempted here.

\begin{figure}
\begin{center}
\includegraphics[width=90mm,clip]{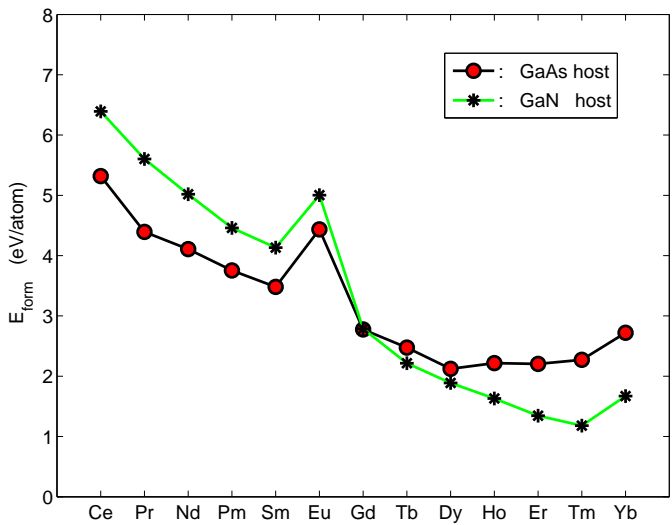}
\caption{
\label{fig:Eform}
(Color online)
Formation energies (cf. Eq. (\ref{EQeform})) 
for substitutional rare-earth ions in GaAs (circles) and GaN (asterix) under 
pnictide-rich conditions (see text for discussions).
For Ga-rich conditions, the formation energies  are larger by 1.2 eV (GaAs) and 2.0 eV (GaN).
}
\end{center}
\end{figure}

The trends of the formation energy of charge neutral substitutional rare-earth ions in GaAs and GaN
are depicted in Fig.   \ref{fig:Eform}. The formation energy 
is defined 
relative to the  pure solids GaAs (GaN) and metallic rare-earth:
\begin{eqnarray}
\label{EQeform}
E_{form}  = 
          E(\mbox{REGa}_{15}\mbox{As}_{16})+\mu_{Ga}- \nonumber \\
    \hspace{25mm}     E(\mbox{Ga}_{16}\mbox{As}_{16})-\mu_{RE},
\end{eqnarray}
where the two   energies, $E(\mbox{REGa}_{15}\mbox{As}_{16})$, 
and $E(\mbox{Ga}_{16}\mbox{As}_{16})$,
 are total energies calculated for the $\mbox{REGa}_{15}\mbox{As}_{16}$ and
$\mbox{Ga}_{16}\mbox{As}_{16}$ supercells. The chemical potential for the rare-earth
is taken as the total energy for the elemental metallic phase 
$\mu_{RE}=E(RE)$, likewise
calculated with the SIC-LSD method.\cite{nature}
$\mu_{Ga}$ is the chemical potential for Ga, which depends on the growth conditions of the
doped semiconductor. During growth the chemical potentials for Ga and As are constrained by
$\mu_{Ga}+\mu_{As}=E(\mbox{GaAs})$. For Ga-rich conditions, excess Ga is assumed
to be in the metallic phase
(orthorhombic $\alpha$-Ga, $Cmca$ symmetry), 
 i.e. $\mu_{Ga}=E(\alpha$-Ga$)$,
while for As-rich (N-rich) conditions excess As (N) 
is assumed to exist as
solid As (rhombohedral $\alpha$-As, symmetry $R\overline{3}m$) 
and $N_2$ gas, respectively, i.e. $\mu_{As}=E(\alpha$-As$)$ and $\mu_{N}=\frac{1}{2}E(N_2)$.
The formation energies in Fig.   \ref{fig:Eform} refer to the pnictide-rich growth conditions. For Ga-rich
conditions, the formation energies of Fig.   \ref{fig:Eform} should be raised by 1.2 eV in the GaAs host and
by 2.0 eV in the GaN host, respectively.
The reference state of the
rare-earth metal is trivalent, except for Eu and Yb, which are divalent, due to the advantage of 
attaining the half-filled $f^7$ configuration, and full $f^{14}$ shell, respectively.\cite{nature}
This causes a relatively high
formation energy of Eu($f^6$) defects in the GaAs and GaN hosts.
Apart from the anomaly of Eu,
the formation energies are seen to exhibit a monotonously decreasing trend through the rare-earth
series with a shallow minimim towards the
later rare-earths Dy-Tm.
The relatively low formation energies are consistent with the high level of doping ($\sim 1 \%$),
which may be obtained for rare-earths in GaN.\cite{steckl}
The formation energies will be somewhat lowered by relaxations of the host
atomic positions in the vicinity of the impurity, which were not considered here.
A check of the relaxation effects was performed for Gd in GaN using the LSD approximation
as implemented in the full-potential version of
the LMTO approach of Ref. \onlinecite{methfessel}.
This calculation revealed  an 11 \% outward relaxation of the 
nearest neighbor N atoms of the Gd impurity, which is in excellent agreement
with the experimental investigations of
Er ions in GaN, finding 10 \% outward relaxation.\cite{citrin}
The use of the LSD approximation for this particular system is justified by the similarity of the
electronic structure of Gd in LSD and SIC-LSD (due to the half-filled $f$-shell), but cannot
be relied upon in general.

A few cases of rare-earth ions substituting for As in GaAs have been considered: Pr, Eu and Tb.
For these, the formation energies are substantially higher, namely 9.0 eV, 8.4 eV, and 8.0 eV,
respectively, again for As-rich conditions (the values are 1.2 eV lower for Ga-rich conditions).
We conclude, that indeed the As-substitution is unfavorable compared to Ga-substitution,
as also evidenced by experiments.\cite{steckl}

\begin{figure}
\begin{center}
\includegraphics[width=90mm,clip]{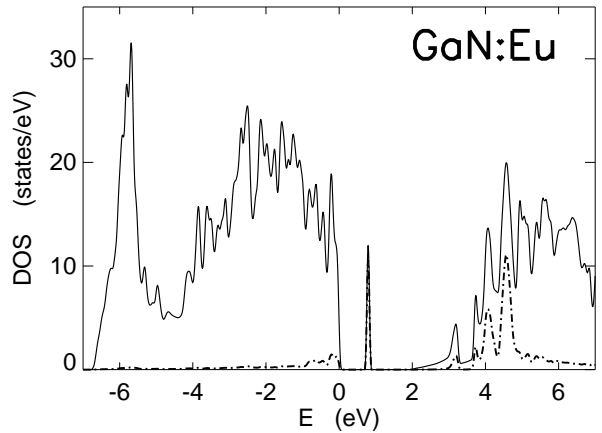}\\
\caption{ 
\label{ganeu}
Electron density of states (DOS)           for Eu in GaN ($x=1/16$).
Full line shows the total DOS of the $x=1/16$ supercell in units of states    per eV and supercell.
The dashed line shows the DOS projected onto the Eu site. The zero of energy is placed at the 
VBM. 
}
\end{center}
\end{figure}

In Fig.   \ref{ganeu} the electronic density of states is illustrated for the case of Eu in GaN. 
The calculated gap is 2.0 eV (in the presence of Eu), defined as the separation of the host
$\Gamma_8$ valence band maximum  state and the 
$\Gamma_6$ conduction band minimum state.
The figure reveals a sharp resonance inside the gap region originating from 
the single Eu majority $f$-state, which is not localized in the $f^6$ configuration, and hence appears as an
unoccupied impurity band state. We discuss this state later,   in subsection III.C. The minority Eu
$f$-states are seen to hybridize into the conduction bands 4-5 eV above the VBM.
The energy position of the localized states are not shown      (see discussion in subsection III.D).

   \subsection{Magnetic Properties}

\begin{figure}
\begin{center}
\includegraphics[width=90mm,clip]{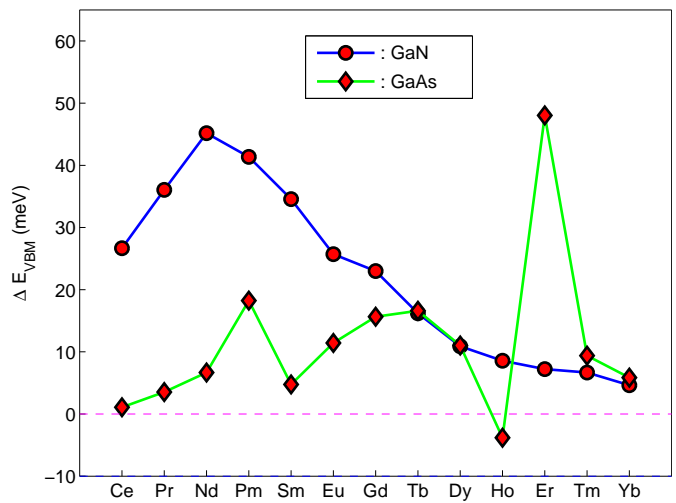}\\
\caption{ 
\label{vbmexc}
(Color online)
Spin-splitting (in meV)
of the electron bands at the valence band maximum
for rare-earth dopants in GaN (circles) and GaAs (diamonds) for $x=1/16$.
A positive sign implies that the band state with its spin
aligned with the rare-earth spin has a lower energy than the
state with its spin anti-parallel to that of the rare-earth ion.
The sign changes of the spin-splitting for GaAs at Ho is caused by
the mixing in of the unoccupied rare-earth impurity $f$-bands into the top
of the valence band.
}
\end{center}
\end{figure}

\begin{figure}
\begin{center}
\includegraphics[width=90mm,clip]{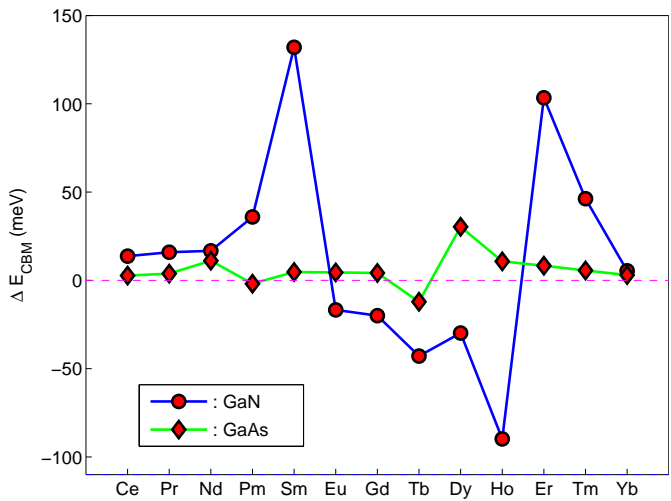}\\
\caption{ 
\label{cbmexc}
(Color online)
Spin-splitting (in meV)
of the electron bands at the conduction band minimum 
for rare-earth dopants in GaN (circles) and GaAs (diamonds) for $x=1/16$.
A positive sign implies that the band state with its spin
aligned with the rare-earth spin has a lower energy than the
state with its spin anti-parallel to that of the rare-earth ion.
The sign changes of the spin-splitting at the conduction band minimum around Sm and Ho
is caused by
the unoccupied rare-earth impurity $f$-bands crossing the conduction edge (cf. Fig.   \ref{bands}).
Note the larger energy scale compared to Fig.   \ref{vbmexc}.
}
\end{center}
\end{figure}

The magnetic interaction of the rare-earth dopants with the host band states is of crucial importance 
for the potential spintronics applications. We analyze this by monitoring 
the exchange splitting, $\Delta E=\epsilon(\downarrow)-\epsilon(\uparrow)$, of the band states at the
valence band maximum (VBM) and the conduction band minimum (CBM), both of which occur at the $\Gamma$ point
in the pure hosts as well as for the $x=1/16$ impurity supercells.
Here the spin-direction $\uparrow$ is taken as the direction of the rare-earth spin (majority spin
direction).
At the VBM,
 the topmost $j=3/2$ fourfold degenerate valence state $\Gamma_8$
splits    into four distinct energies separated by a few meV. Two of these are essentially 
pure spin-up and spin-down, respectively,
(corresponding to the $m_j=\pm 3/2$ quantum numbers), and Fig.   \ref{vbmexc} 
shows the calculated differences in energy between these two states.
The CBM state is the twofold $\Gamma_6$ state, which is antibonding $s$-like, and its splitting
 in the presence of magnetic rare-earth ions is depicted in Fig.   \ref{cbmexc}

\begin{figure}
\begin{center}
\includegraphics[width=90mm,clip]{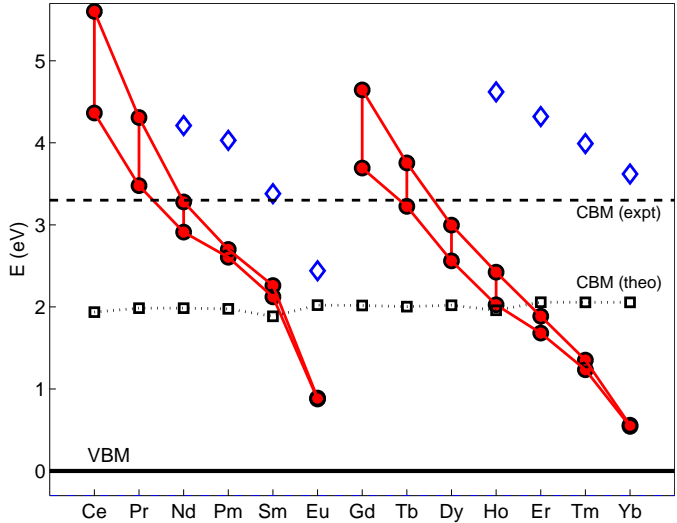}\\
\caption{ 
\label{bands}
(Color online)
Trends in energy band structure around the fundamental gap for 
RE$_x$Ga$_{1-x}$N, $x=1/16$, supercells. The VBM is at zero energy, full circles mark the unoccupied $f$-bands,
while the squares and dotted line mark the calculated CBM. The $f$-band is defined 
by the smallest and largest $f$-dominated eigenvalue at the $\Gamma$-point (considering
majority spin only, for first
half of the rare-earth-series), and its width is mainly due to spin-orbit
splitting.   The experimental gap (3.3 eV) is marked
with a dashed line. The calculated gap (for pure GaN) is 1.77 eV.
The diamonds mark the calculated positions of the $\epsilon(0/-)$ levels in the localized picture, cf.
Table I and subsection III.C.
}
\end{center}
\end{figure}

The spin-splitting of the top of the valence band in GaN is seen to peak around Nd at a value of 
45 meV (Fig.   \ref{vbmexc}), and falling smoothly towards a value of 6 meV in Yb. In contrast,
the spin-splitting of the valence band of GaAs fluctuates more, which is due to the interaction with the
unoccupied $f$-states. The latter appear in the SIC-LSD description as impurity bands, moving
down with respect to the host bands, as one moves across the first half of the lanthanide series,
and again similarly for
the second half of the series. This is illustrated for GaN in Fig.   \ref{bands}.
The hybridisation effect of the $f$-bands with the VBM $p$-bands, as well as with the CBM $s$-bands can be 
substantial,\cite{dalpian} if the $f$-bands are close in energy to the gap  edges. 
In the case of GaAs the $f$-impurity band actually merges with the
VBM for Sm and Eu dopants, while  for Pm dopants the $f$-band is just above the VBM.  
Due to level repulsion the majority spin
$p$-band at the VBM for GaAs:Pm is pushed  significantly more
down. Similarly, for the other  spin channel in the second half of the
lanthanide series, the
$f$-band crosses the VBM between Ho and Er, and for          Yb the $f$-like eigenvalue at the
$\Gamma$-point is below the
4-fold degenerate host VBM states (but only at $\Gamma$ - the impurity band is about one third filled at self-consistency).

The sign of the spin-splitting 
at the top of the valence band
is in all cases, except Ho in GaAs, positive, i.e. the VBM $p$-states tend to align with the
rare-earth spin. Holes
will therefore predominantly be created in the minority spin channel, i.e. the $p$-type carriers will
be spin-aligned with the rare-earth dopant.
A similar small interaction was found for Gd in GaN in Ref.
\onlinecite{dalpian}, which did not include spin-orbit interaction. 
However, in that work the opposite sign of the spin-splitting was reported.
The origin of this sign difference
most probably lies in the different treatment of the occupied Gd $f$-states. In 
Ref. \onlinecite{dalpian} the $f$-states are treated as band states, which fall inside the
occupied valence bands and interact strongly with the 
valence band top states. The repulsion effect on the majority spin states at the VBM is so strong, that
these are pushed above the minority VBM state, leading to a sign change of $\Delta E$.

In contrast to the rather modest spin-splitting obtained for the rare-earth dopants in the present
work, for Mn in GaN and GaAs it is much stronger.
The spin-splitting induced by the Mn localized spins has been calculated\cite{NatMat}
by the SIC-LSD method  to be $\Delta E= -0.55$ eV and 
$-0.43$ eV in GaN and GaAs (for $x=1/16$), respectively, {\it i.e.}
one or two orders of magnitude larger (and of opposite sign) than the splittings shown in Fig.   
\ref{vbmexc}. 
For Mn in GaAs and GaN, the overlap of the Mn $d$-states (in particular those of $t_2$ symmetry) with the
host states at the VBM is significantly larger than that of the rare-earth ions. Hence, for a Mn impurity
with localized bonding and Mn $d$ dominated states, the highest
state at the VBM becomes the antibonding host-$t_{2}$-Mn $d$ band-like resonance, which is pushed up.
At the same time, in the minority channel a bonding host-$t_{2}$-Mn $d$ resonance is formed and moving down
in energy, altogether leading to a negative  spin-splitting 
$\Delta E=\epsilon(\downarrow)-\epsilon(\uparrow)$.\cite{NatMat}
The calculations reported in Ref. \onlinecite{NatMat} did not include the spin-orbit interaction,
but the conclusions of that work prevail when spin-orbit interaction is included.\cite{unpub}
In the case of rare-earth dopants the $f$ orbitals are much better confined to the rare-earth site and 
their hybridizations are significantly weaker, so that the spin-splitting becomes dominated by the direct
exchange between host-states and rare-earth site (described through the spin-dependent effective potential).
The only exception occurs when the unoccupied $f$-bands accidentally fall close to the host band edges, as discussed.
Hence we conclude that  for a $p$-type GaAs or GaN host there will
be a much weaker spin polarization effect induced on the hole carriers by
rare-earth dopants than when transition metal dopants are used.

 GaN is most often $n$-type, so it is 
more appropriate to investigate the rare-earth induced effects on the conduction band states.
This is a more difficult matter, since the LDA bands of the unoccupied conduction states
are known to be inaccurate. Most importantly, the fundamental gaps are seriously
underestimated, but in some cases the entire shape of the conduction bands is in error, 
{\it e.g.}
in GaAs.\cite{nec} In the present approximation, the GaAs and GaN band gaps are calculated
to be 0.11 eV and 1.77 eV, respectively, as compared to the experimental values of
1.42 eV\cite{blakemore} and around 3.3 eV\cite{iza2}, and to full-potential LMTO values of
0.21 eV\cite{iza} (see also Ref. \onlinecite{gbb}) and 2.2 eV.\cite{iza2}
A further complication is the appearance of the unoccupied $f$-bands of the
rare-earth dopant, which interact with the host conduction band states, in particular with the
$s$-like CBM.\cite{dalpian}
The physical meaning     of these $f$-bands depends on whether the band picture applies for
an $f$-like electron added onto the rare-earth ion, which is doubtful.     For example,
the  observation of distinct multiplet lines in the photoluminescence spectra of rare-earth
doped GaN\cite{steckl} clearly demonstrates the localized nature of the $f$-electrons.
See also the discussion in the next subsection.

The spin-splitting of the CBM states seen in Fig.   \ref{cbmexc} 
is minute in GaAs, but much stronger in 
GaN, in particular where the unoccupied $f$-bands cross the CBM, which occurs around Sm and Ho (see
Fig. \ref{bands}).
In the mid-lanthanide series, for Eu, Gd, Tb, Dy, and Ho,
the exchange interaction is antiferromagnetic, i.e. the minority
spin CBM state is lower than the majority spin state. This is again caused by the level repulsion effect.
In the case of
Eu, the unoccupied band of the seventh  majority spin electron has moved below the conduction
band edge thus pushing the majority spin $s$-state at CBM up. For Gd, Tb, Dy and Ho, the unoccupied
$f$-band states are of minority spin, and situated above the CBM, thus pushing the minority spin
$s$-like CBM state down below the majority spin $s$-like CBM state.
This behavior was also demonstrated for Gd in hypothetical zinc-blende GdN in Ref. \onlinecite{dalpian}. 

\begin{figure}
\begin{center}
\includegraphics[width=90mm,clip]{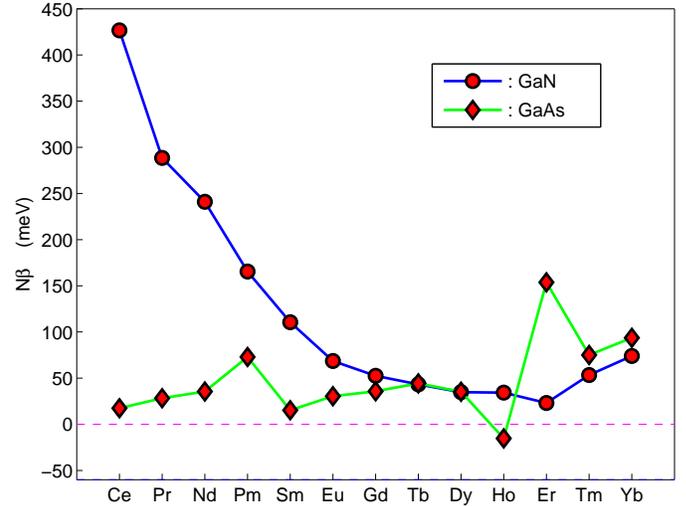}\\
\caption{ 
\label{beta}
(Color online)
Exchange interaction parameter $N\beta$
of the valence band maximum
for rare-earth dopants in GaN (circles) and GaAs (diamonds) for $x=1/16$.
}
\end{center}
\end{figure}

\begin{figure}
\begin{center}
\includegraphics[width=90mm,clip]{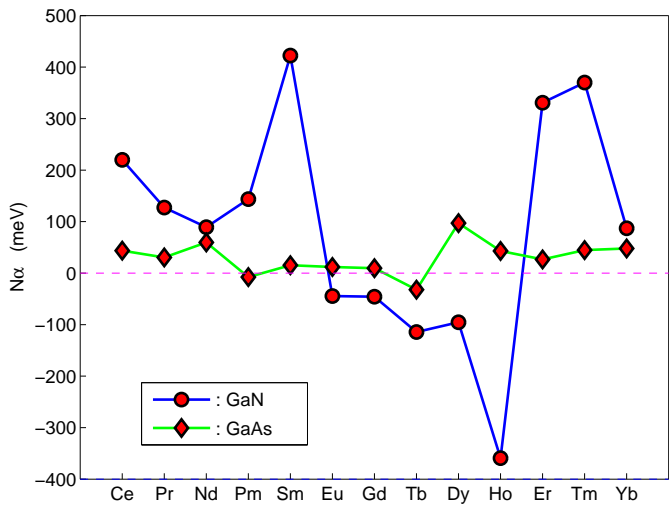}\\
\caption{ 
\label{alfa}
(Color online)
Exchange interaction parameter $N\alpha$
of the conduction band minimum 
for rare-earth dopants in GaN (circles) and GaAs (diamonds) for $x=1/16$.
}
\end{center}
\end{figure}

To further analyze the exchange
interaction of the rare-earth ion with the $p$-like VBM, the relevant  exchange integral, 
$J_{pf}$
is  extracted using the expression:\cite{larson}
\begin{equation}
N\beta\equiv NJ_{pf}=\frac{\Delta E_{VBM}}{2Sx}.
\end{equation}
Similarly, the exchange-splitting of the CBM $s$-like state can be quantified through an 
$s-f$ interaction parameter:
\begin{equation}
N\alpha\equiv NJ_{sf}=\frac{\Delta E_{CBM}}{2Sx},
\end{equation}
In these expressions, $S$ is the total spin of the rare-earth ion, $x$ the dopant concentration, and
$N$ the
number of host formula units within the
normalization volume entering the definition of exchange integrals.\cite{mahan}
Figures \ref{beta} and \ref{alfa} show the values of these parameters through the rare-earth series.

The exchange integral parameters are seen to reflect the behaviors already discussed for the
spin-splittings in Figs.   \ref{vbmexc} and \ref{cbmexc}.
For Gd in GaN we may compare to the exchange parameters extracted in Ref. \onlinecite{dalpian} for GdN,
whose estimated coupling parameters $N\beta=-0.01$ eV and $N\alpha=-0.4$ eV are showing
somewhat weaker coupling to the VBM but significantly stronger coupling to the CBM, as compared to
our respective values  of $N\beta= 53$ meV and $N\alpha=-46$ meV, calculated  
for the diluted Gd in GaN (see Figs.  \ref{beta} and \ref{alfa}).

In contrast to the rather modest interaction parameters obtained for the rare-earth dopants here, 
for dilute Mn in GaN and GaAs, 
the exchange interaction of the Mn localized spins with the VBM  was
obtained as  $N\beta=   -2.5 $ eV in GaN and  $N\beta=   -1.5 $ eV in GaAs.\cite{NatMat}

In conclusion, it appears very unlikely that isolated 
rare-earth dopants by themselves can cause
the room temperature ferromagnetism, let alone the giant magnetization amplifications, 
observed in the experiments by Ref. \onlinecite{ploog}.

\subsection{Acceptor levels}

The behavior elucidated in Figs.   \ref{cbmexc} and \ref{alfa} can only be considered a trend study,
since the absolute positions of the conduction bands are not accurate. The LSD is known
to substantially underestimate the band gap of semiconductors.\cite{perdewlevy}
 In the SIC-LSD approach the host band
states are described by the LSD approximation, however  applied to the self-consistent charge density
as obtained by the minimization of the SIC-LSD energy functional, Eq. \ref{Esic}, which deviates  somewhat
from the self-consistent charge density obtained by minimizing the LSD total energy functional.
Generally, the SIC-LSD leads to larger magnetic
moments and larger $f$-electron occupation than the LSD,
although sometimes the difference can be rather small. 
Hence we may anticipate that the gap problem of LSD prevails in
SIC-LSD.

Even more problematic is the 
description of the unoccupied $f$-states as weakly correlated LSD bands, which is in conflict with
the localized picture used for the occupied part of the $f$-manifold. The LSD potential is derived from
the limit of nearly-free electrons, i.e. fast electrons moving in        an effective
potential as given by the LSD approximation. This is a good approximation for most cases, 
{\it e.g.} for the
occupied states of the host materials, which are built from extended $s$- and $p$-orbitals of the constituent
atoms. The approximation is less adequate for the spatially confined $f$-orbitals of the rare-earth
dopant, for which the overlap with neighbouring impurity atoms is small. Hence, an extra
added  $f$-electron would 
tend to stay for a long time 
on one rare-earth ion before hopping to a neighbour. The local environment will have
time to
adjust to the presence of the extra electron, and the effective potential will be modified.
The same effect is well known for localized gap states in semiconductors, for which the transition state
approximation is often invoked, i.e. the defect energy level is calculated by 
occupying the defect state with half of an electron.
In this subsection we will investigate various approaches to the calculation of the $f$-related 
acceptor levels $\epsilon(0/-)$ 
corresponding to the addition of one $f$-electron to the $f$-shell.

In the delocalized limit, as discussed, the added $f$-electron is put in the lowest unoccupied 
$f$-impurity band, which is positioned at
\begin{equation}
\label{epsLSD}
\epsilon_{LSD}=<f_{n+1}|H_{LSD}|f_{n+1}>,
\end{equation}
where $|f_{n+1}>$ is the one-particle $f$-state, and $H_{LSD}$ is the LSD Hamiltonian evaluated with
the self-consistent charge density as obtained by minimization of the $E^{SIC-LSD}$ functional. 
Generally, several unoccupied $f$-states exist, 
and the lowest value of the above matrix element must be taken to represent the minimal addition energy.

In the opposite limit of localized
$f$ states one may calculate the position of the $\epsilon(0/-)$ level of a divalent rare-earth ion with
one additional localized $f$-electron, by an SIC-LSD total energy difference:
\begin{equation}
\label{Eneg}
\epsilon(0/-)=E(f^{n+1},-)-E(f^n,0)
\end{equation}
Here, $E(f^n,0)$ is the total energy of the 
charge neutral trivalent rare-earth ion, while 
$E(f^{n+1},-)$ refers to the total energy of a divalent rare-earth ion with an extra
electron in the supercell, and a homogeneous positive background charge of $+e$ added.
In the approximation implied by Eq. (\ref{Eneg}) 
it is thus assumed that the presence of an extra electron on the rare-earth ion may be artificially
compensated by the positive background charge, rendering the supercell altogether electrically neutral,
which is necessary in order to have a well defined total energy minimization problem.
Obviously, spurious electrostatic interactions are introduced by this approach, 
and the reliability of the scheme   must be checked.
The same approach is often invoked for other charged impurities in 
semiconductors.\cite{walle,semi}

A good test  of the approximation in Eq. (\ref{Eneg}) is provided by the charge neutral 
SIC-LSD approximation for the $f$ addition level.
The latter is defined as the SIC-LSD total energy difference between
divalent and trivalent rare-earth ions  in a charge neutral supercell, {\it i.e.} the
extra electron added to the rare-earth is simply supplied from the highest valence band state.
This corresponds to the energy differences displayed by the $E(II)-E(III)$ curve in Figs.   \ref{GaAs_val}
and \ref{GaN_val}:
\begin{equation}
\label{Eneg2}
\epsilon(0/-)_{appr}-\epsilon_{VBM}=E(f^{n+1},0)-E(f^n,0).
\end{equation}
In this approximation the net accumulation of negative charge in the vicinity of the rare-earth ion is
simply supplied from the Fermi level, however due to the finite supercell, uncontrollable 
interactions between the
negative $f$-ion and the hole state introduced will occur. The charge density of the
host hole state does not correspond to a completely homogeneous charge density, and furthermore, the
missing electron charge influences also the exchange-correlation part of the energy functional. Hence the
two approaches differ, but as we will show give quite similar results, so that we can conclude that indeed the
Coulombic interaction caused by the finite supercell is  small.

 Finally, we can also improve on the delocalized description of the $f$ acceptor level by 
considering the transition state. The position of the lowest unoccupied $f$-band at
half occupancy may be approximated 
as the average of the position of this band at zero and full occupancy:
\begin{equation}
\label{epsTS}
\epsilon_{TS}^+=\frac{1}{2}(\epsilon_{LSD}(II)+\epsilon_{LSD}(III)).
\end{equation}
At zero occupancy, $\epsilon_{LSD}(III) $ 
corresponds to Eq. (\ref{epsLSD}), and at full occupancy
$\epsilon_{LSD}(II)$ 
corresponds to the same expression, however
evaluated with the LSD Hamiltonian for the self-consistent charge density corresponding to
the divalent rare-earth configuration.
In this expression, $|f_{n+1}>$ must be that particular $f$-state, which is delocalized 
in the trivalent but localized in the divalent configuration.
The difference between the two LSD eigenvalues entering the average can be identified as
an effective Coulomb $U$ parameter, and
the transition state energy is therefore equivalent to a shift by $\frac{1}{2}U$ of the LSD
unoccupied band.

\begin{table}[h]
\label{level}
\begin{ruledtabular}
\begin{tabular}{|c|c|cccc|}
 Host/gap   & dopant  &  \multicolumn{1}{c}{$\epsilon(0/-)$}  & 
                     \multicolumn{1}{c}{$\epsilon(0/-)_{appr}$} & 
                     \multicolumn{1}{c}{$\epsilon_{LSD}$}       & 
                     \multicolumn{1}{c|}{$\epsilon_{TS}^+$}                    \\
 \hline
     & Pm &        2.30  &  2.13  &  0.37  &  2.39  \\
{\bf GaAs} & Sm &        1.47  &  1.28  &  0.13  &  1.96  \\
     & Eu &        0.89  &  0.80  &  0.00  &  1.88  \\
expt:& Gd &        4.17  &  4.20  &  1.71  &  4.98  \\
 1.42& Tb &        3.25  &  3.23  &  0.61  &  3.13  \\
     & Ho &        2.13  &  2.15  &  0.24  &  2.83  \\
theo:& Er &        1.85  &  1.70  &  0.00  &  2.59  \\
 0.11& Tm &        1.12  &  1.20  &  0.00  &  2.41  \\
     & Yb &        0.89  &  0.97  &  0.00  &  2.38  \\
\hline
     & Nd &        4.21  &  4.43  &  2.91  &  5.09  \\
{\bf GaN} & Pm &        4.03  &  4.06  &  2.56  &  4.90  \\
     & Sm &        3.38  &  3.11  &  2.12  &  3.69  \\
expt:& Eu &        2.44  &  2.15  &  0.61  &  2.79  \\
3.3  & Ho &        4.62  &  4.44  &  1.99  &  4.57  \\
     & Er &        4.32  &  4.19  &  1.65  &  4.44  \\
theo:& Tm &        3.99  &  3.89  &  1.17  &  4.37  \\
1.77 & Yb &        3.62  &  3.45  &  0.57  &  3.78  
\end{tabular}
\end{ruledtabular}
\caption{Calculated defect levels of divalent rare-earth dopants in GaAs and GaN. First column gives the
host with experimental and theoretical
gap values, second column the dopant, third column the calculated defect level on the basis
of the SIC-LSD total energy difference, Eq. (\ref{Eneg}), fourth column the same with the approximate expression,
Eq. (\ref{Eneg2}), i.e. neglecting the charging of the near defect region, fifth column the LSD position of the
lowest unoccupied $f$-band, Eq. (\ref{epsLSD}), and sixth coulmn the transition state as defined in 
Eq. (\ref{epsTS}).
All numbers are in eV and relative to the VBM.
}
\end{table}

In Table I we compare for selected cases
the $f$-related $\epsilon(0/-)$ level in Eq. (\ref{Eneg}) with the approximate expressions 
given by equations (\ref{epsLSD}), (\ref{Eneg2}) and (\ref{epsTS}). The theoretical and
experimental host gaps are quoted for comparison. The most consistent interpretation of the 
present results is to compare the calculated levels with the theoretical band gap, but often
comparisons to the experimental gap in fact are done. After all, the band gap
discontinuity\cite{perdewlevy} problem persists in the expressions (\ref{Eneg})-(\ref{epsTS}). 
One observes that in all cases the 
negatively charged rare-earth impurity has a higher energy than the theoretical band gap, and hence
appears as a scattering resonance in the
conduction bands.  
The implication is that even in $n$-type material the rare-earth ions will remain trivalent.
If compared to the experimental band gap, only
for Eu, Tm and Yb in GaAs and for Eu in GaN
does the $\epsilon(0/-)$ 
level fall below the experimental conduction band edge, while for 
Sm the $\epsilon(0/-)$ level comes rather close to the experimental conduction edge.
Photoluminescence of (wurtzite) GaN:Eu thin films reveals  no
signature of divalent Eu ions,\cite{nyein} thus corroborating the conclusions drawn by comparison to 
the theoretical gap.

Through the rare-earth series, the
approximate expression in Eq. (\ref{Eneg2}) differs from Eq. (\ref{Eneg}) by only 
$0.1-0.3$ eV, which is reassuring, since it shows that the spurious
charge interaction  effects of the divalent 
rare-earth ion in the finite supercell 
are quite small. The correction with respect to the LSD band position is significant,
generally several eV, 
{\it i.e.} the local Coulombic corrections due to the presence of the added localized
$f$-electron are substantial. 
As explained, the transition state construction seeks to
take this into account, and indeed brings the level position in much better agreement
with the expression in Eq. (\ref{Eneg}),  although it tends to mostly overshoot, in some cases by more
than 1 eV.

\subsection{f removal energies}

The calculation of the energy position of the localized states in SIC-LSD theory has always been a
matter of concern. The $f$-removal
energies one observes in photoemission include all the atomic multiplets of the $f^{n-1}$ ion
left behind, which can lead to a spectral function with $f$-related features over a $\sim 10$ eV range,
see {\it e.g.}  Ref. \onlinecite{seb}. Even calculating the energy of the most stable state with one 
$f$-electron removed is difficult. First of all, this will usually not be the ground state for an 
$N-1$ electron system, so applying a density functional theory based formalism poses a problem. 
But also in practice, the
self-consistency process for a rare-earth with a hole created in the $f$ shell
leads to unwanted effects, since the enhanced Coulomb attraction due to the missing $f$-electron 
will pull down the unoccupied $f$-levels below the Fermi level. Hence, if the initial state corresponds
to a localized $f^n$ configuration, the effective rare-earth configuration in the final state
will not be $f^{n-1}$, but somthing in between $f^{n-1}$ and $f^{n}$, often closer to the latter, and hence
a too small removal energy is obtained (photoemission experiments do interprete certain features in terms
of such 'well screened' peaks\cite{pes-f}). On the other hand, just knocking out 
the $f$-electron but not driving the charge density to self-consistency overestimates the $f$ removal energy
by not allowing for screening of the charge hole
by non-$f$ electrons. What is physically relevant is an intermediate picture, where
the fast non-$f$ electrons are allowed to screen the charge hole, while the $f$-electrons are not allowed to
respond to the removed charge, a situation which is not easily implemented in a total energy scheme.   

 According to  Janak's theorem,\cite{janak} the derivative of the LSD energy functional with respect to
orbital occupation is equal to the LSD eigenvalue for the corresponding orbital, which can be taken as a good 
representative of the removal energy of an electron in that state, if the charge of the orbital is everywhere 
small, as it will be for a band-like state. A similar theorem holds in SIC-LSD,\cite{pederson} where the
SIC eigenvalue
\begin{equation}
\label{epsSIC}
\epsilon_{SIC}=<f|H_{LSD}+V_{SIC}|f>
\end{equation}
represents the derivative of the SIC-LSD total energy with respect to the occupation of the 'canonical orbitals',
which in the context of a periodic solid means a Bloch state formed from the localized $f$-states on
all the rare-earth sites. Here, $V_{SIC}$ represents the aditional potential term arising in the
self-consistency equation due to the self-interaction term in Eq. (\ref{dsic}).
Hence, using the SIC eigenvalues, the
occupied $f$-states (without multiplets) appear as sharp resonances (SIC-LSD 'bands'\cite{pederson})
 below the valence bands (at around $-11.5$ eV for the case of Eu in GaN depicted in Fig.  \ref{ganeu}), 
which is unrealistically deep for comparison to physical removal energies as observed in photoemission.
The reason is that photoemission knocks out a localized $f$-electron rather than a 'canonical' $f$-state.
 
Similar to the previous subsection, we may however apply
a physically more reasonable but theoretically less rigorous approach by
placing the removal energies of $f$-states at the transition state position. The eigenvalue given by
Eq. (\ref{epsSIC}) represents the energy cost due to removal of the first infinitesimal part of the
$f$-electron (since  $V_{SIC}$ is evaluated for the initial ground state, {\it i.e.} at full occupancy), but as
more and more of the $f$-electron is removed, the SIC potential term decreases, and eventually, only $H_{LSD}$
is left. As an average we may therefore take the $f$ removal energy as midway between the 
SIC-LSD and LSD band positions:
\begin{equation}
\label{epsTS2}
\epsilon_{TS}^-=\frac{1}{2}(<f|H_{LSD}+V_{SIC}|f>+<f|H_{LSD}|f>).
\end{equation}
In effect, the SIC potential is only counted with half its strength in the transition state approximation to
the removal energy.
By evaluating $H_{LSD}$ in the initial state, {\it i.e.} without the hole in the $f$ shell,
 we avoid the aforementioned 
effect of the $f$-hole pulling the $f$ levels down. 
The transition state philosophy was also implemented in
Ref. \onlinecite{spaldin}, albeit in a different manner, by invoking the averaging factor of $\frac{1}{2}$ already
in the total energy functional, while we do it here only for the removal energy, Eq. (\ref{epsTS2}), 
after self-consistency.

\begin{figure}
\begin{center}
\includegraphics[width=90mm,clip]{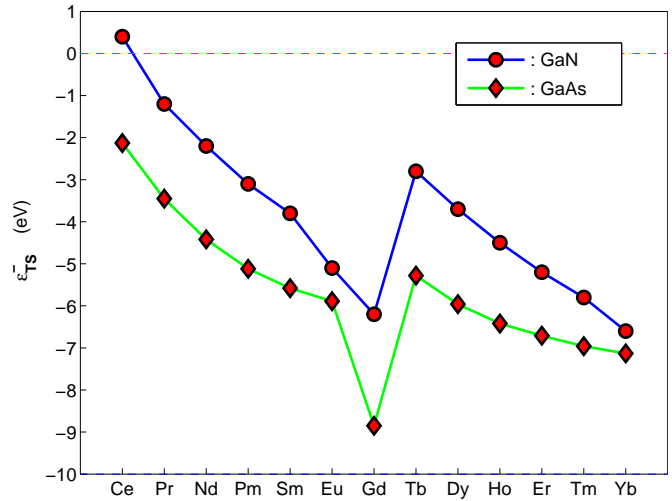}\\
\caption{ 
\label{Fminus}
(Color online)
Transition state evaluation of the $f$-removal energies in GaAs (diamonds) and GaN (solid circles). The
units are eV and relative to the VBM.}
\end{center}
\end{figure}

Figure \ref{Fminus} displays the trends of 
$ \epsilon_{TS}^-$ through the lanthanide series in both GaAs and GaN.
The gradual increasing binding of the $f$-shell is clearly reflected in the transition state, only broken at Gd
in the middle of the series. 
Furthermore, the $f$ removal energy is consistently smaller (less negative) in GaN compared to GaAs.
A few  experimental data corroborate the interpretation of the transition state as
a good measure of the $f$ removal energy: In ref. \onlinecite{Maruyama}, the photoemission spectra of
Eu doped GaN reveal the signature of the $f^6\rightarrow f^5$ transition in the interval between
5 and 10 eV below
the Fermi edge, {\it i.e.} the lowest $f^5(^6H)$ multiplet is situated at -5 eV, in excellent agreement with
the value -5.1 eV in
Fig.   \ref{Fminus}. For GdN films, the Gd $f$-emission peaks at -7.8 eV\cite{leuenberger}, compared to the
transition state energy of dilute Gd in GaN of -6.2 eV in Fig.   \ref{Fminus}. For YbN, the similar
onset of $f$-emission is estimated to occur
 around -6 eV,\cite{wachter} compared to the value -7.1 in Fig.   \ref{Fminus}.
In CeAs, $f$-emission occurs around -2.5 eV,\cite{campagna} as compared to the transition state estimate of
-2.1 eV for dilute Ce in GaAs. In GdAs, $f$-emission occurs at -9 eV,\cite{yamada} while the estimate of
Fig.   \ref{Fminus} for Gd in GaAs is -8.9 eV.
 For ErAs, the 
$f$-emission starts at -5 eV,\cite{komesu} while the transition state position of dilute Er in GaAs is
calculated as -6.7 eV.

\section{Summary}    

Trends in the electronic structure of rare-earth substitutional impurities in GaAs and cubic GaN
have been discussed based on self-interaction corrected local spin density calculations. The
trivalent configuration is found as the ground state of most dopants, while the divalent acceptor
level is found above the theoretical band gap, and within the 
experimental gap only for Eu, Tm and Yb in GaAs and for Eu in GaN. The magnetic interaction
of the rare earth ion with the host states at the valence and conduction band edges has been 
investigated and found to be relatively weak in comparison with $3d$ impurities. 
Hence it is unlikely that the rare-earth dopant by itself 
may induce room-temperature ferromagnetism and gigantic magnetic enhancement, as observed in certain
experiments.\cite{ploog} To explain these experiments it seems imperative to include     the interaction of the 
rare-earth dopants with other defects, native or external, which will be pursued in our future research.

\section{acknowledgements}

This work was partially funded by the EU Research Training Network
(contract:HPRN-CT-2002-00295) 'Ab-initio Computation of Electronic 
Properties of $f$-electron Materials'. AS and NEC acknowledge support from the Danish Center for Scientific
Computing.  Work of LP was
sponsored by the Office of Basic Energy Sciences, U.S. Department of Energy.

\end{document}